\documentstyle[12pt, amssymb, epsfig,graphicx
]{article}

\newcommand{\eeq}{\end{equation}}
\newcommand{\beq}{\begin{equation}}
\newcommand\slabel[1]{\label{#1}}

\newcommand\eq[1]{Eq.~(\ref{#1})}
\newcommand\eqs[2]{Eqs.~(\ref{#1}) and (\ref{#2})}

\newcommand{\AmS}{{\protect\the\textfont2
  A\kern-.1667em\lower.5ex\hbox{M}\kern-.125emS}}

\begin{document}

\thispagestyle{empty}
\vspace*{.5cm}
\noindent
%\today
%\hfill {\tt hep-ph-ph/0502xxx}\\
\vspace*{1cm}
\begin{center}
{\Large\bf A Separate Universe Approach to Quintessence Perturbations
}\\[.8cm] {\large Christopher Gordon 
}\\[.6cm]
{\it   Kavli Institute for Cosmological Physics, Enrico Fermi
  Institute and  Department of Astronomy and Astrophysics,
University of Chicago, Chicago IL 60637
}
\end{center}
\vskip 1 cm
%\date{\today}
%\maketitle

\begin{abstract}
There is some observational evidence that the dark energy may not be
smooth on large scales. This makes it worth while to try and get as
simple and as intuitive a picture of how dark energy perturbations
behave so as to be able to better constrain possible models of dark
energy and the generation of large scale perturbations. The separate
Universe method provides an easy way to evaluate cosmological
perturbations, as all that is  required is an understanding of the background
behavior. Here, this method is used to show how the size of the
dark energy perturbations, preferred by observations, is larger than
would be expected, and so some mechanism may be required to amplify them.   
\vspace{1pc}
\end{abstract}

%\maketitle

\section{Introduction}
One way to rule out a cosmological constant would be to show that the dark
energy in not spatially homogenous. Recently, it was found that
the low quadrupole in the WMAP data \cite{Ben03,Spe03,lowquad} favors
perturbations in the dark energy at the two sigma level
\cite{MorTak03,GorHu04}. The dark energy was taken
to be a canonical scalar field which is usually referred to 
as the `quintessence'. The perturbations were evaluated using the
metric approach \cite{Bar80}. This requires specifying a foliation
of the space-time into constant time hypersurfaces or equivalently
the choice of a gauge. In reference \cite{MorTak03}, the Newtonian gauge
is used and in reference \cite{GorHu04} the comoving gauge.

An alternative way of treating perturbations in cosmology is the
`separate Universe' approach, in which the background equations can be
perturbed directly  to get the large scale perturbations
in the flat gauge \cite{sepuni}. This has the advantage that the
derivation and solution of the perturbation equations follow simply
from the background equations. It also provides a more intuitive
picture of how the perturbations behave.

In this article we study the quintessence perturbations using the
separate Universe approach. There are many other treatments of
quintessence perturbations, for example in the Newtonian gauge
\cite{BarCorLidMal04}, in the comoving gauge \cite{GorHu04} and in a
gauge invariant formulism \cite{MalWan05}. I hope the treatment in
this article will be complementary to those, and help to provide a
clearer picture on the nature of density perturbations in
quintessence.

\section{Separate Universe Method}
\slabel{sepunisec}
This method provides a simple way of modeling large scale
inhomogenities.  Here we give a pedagological explanation, which
emphasizes its casual underpinnings. See  references \cite{sepuni}
for more detailed derivations.

If the Universe is spatially smooth and the overall geometry is flat, then Friedmann equation relates the scale of the Universe, $a$, to its overall density, $\rho$,\beq
H^2 \equiv \left( {{\rm d}a \over {\rm d}t}{1\over a} \right)^2 = {1 \over 3M_p^2} \rho
\slabel{friedmann}
\eeq
where $H$ is known as the Hubble parameter,
$t$ is the time and $M_p\equiv (8\pi G)^{-1/2}=2.436\times 10^{18}{\rm
  GeV}$ is the reduced Planck mass. The pressure, $p$, is related to
the density by the equation of state parameter 
\beq
w \equiv p / \rho.
\slabel{w}
\eeq
It has the value of zero for non-relativistic matter and a third for
radiation. For a canonical scalar field it is greater than or equal to
minus one. Unless otherwise specified, $w$ will be assumed to satisfy
\beq
|w| < 1.
\eeq
Einstein's theory of general relativity also relates the acceleration
of the scale factor to the matter density and pressure
\beq
{{\rm d}^2 a \over {\rm d} t^2} = -{1 \over 6 M_p^2}(\rho + 3p)a.
\slabel{acceleration}
\eeq
The continuity equation is given by
\beq
{{\rm d}\rho\over {\rm d}t} + 3H(1+w)\rho=0.
\slabel{continuity}
\eeq
Assuming the equation of state, $w$, is constant, the solution to the above
equation is
\beq
\rho = {\rho_0 \over a^{3(1+w)}}
\slabel{density}
\eeq
where $\rho_0$ is the density at $a=1$.
The distance a signal, traveling at the speed of light, can travel is
\beq
D \equiv a \int_{t_i}^{t_f} {1  \over a} \, {\rm d}t
\slabel{distance}
\eeq
where the units are chosen so that the speed of light is one, and $t_i$
and $t_f$ are the initial time and final times respectively.
It is convenient to express time in terms of the number of efolds of expansion
\beq
N \equiv\log(a).
\slabel{N}
\eeq
Then from the definition of the Hubble parameter
\beq
{dN \over dt} = H
\slabel{Ntimerateofchange}
\eeq
which when substituted into the distance equation (\eq{distance}) gives
\beq
D= a \int_0^N {1\over a} {1 \over H} \, dN.
\slabel{distance1}
\eeq
Using the equation for the density (\eq{density}) and the definition
of efolds (\eq{N}), the Friedmann equation (\eq{friedmann}) can be
expressed as  
\beq
H^2 = {1 \over 3 M_p^2} {\rm e}^{-3(1+w)N}.
\slabel{friedmann1}
\eeq
Substituting this into \eq{distance1}  and solving the integral gives
\beq
D = {2\sqrt{3} \over 1+3w }{M_p \over \sqrt{\rho_0}} {\rm e}^N \left({\rm
  e}^{N(1+3w)/2} -1 \right).
\slabel{distance2}
\eeq
The Hubble distance is defined as 
\beq
D_H \equiv {1 \over H}. 
\slabel{hubbledistance}
\eeq
Using the distance equation (\eq{distance2}) and the Friedmann equation
(\eq{friedmann1}) gives
\beq
{D \over D_H} = {2 \over {1+3w}}\left(1 - {\rm e}^{-{1\over
    2}(1+3w)N} \right).
\slabel{doverdh}
\eeq
From the Friedmann equation (\eq{friedmann1}), the distance
equation (\eq{distance2}) and the definition of 
efolds (\eq{N}) 
\beq
{D \over a D_H|_{N=0} }= {2 \over (1+3w)} \left(
{\rm e}^{{1\over 2}(1+3w)N}-1\right).
\slabel{dovera}
\eeq
As can be seen from \eqs{doverdh}{dovera}, the value of $w=-1/3$ is
special. For $w<-1/3$, the acceleration equation (\eq{acceleration})
and the definition of the equation of state (\eq{w}) give an
accelerating scale factor. This is thought to have occurred in the
early Universe during a period known as inflation \cite{infl}. For
large $N$ and $w<-1/3$, the ratio of the red shifting initial Hubble parameter to
the casual distance (\eq{dovera}) tends to 
\beq
{D \over a D_H|_{N=0} } \rightarrow {-2 \over 1+3w}.
\slabel{dovera1}
\eeq
It follows that, points that are initially more than of order a Hubble distance apart are
always out of causal contact as long as inflation lasts.

After inflation, there is the radiation dominated era with $w=1/3$
followed by the matter dominated era with $w=0$. When $w>-1/3$, the
ratio of the casual distance to the Hubble distance (\eq{doverdh}) 
tends to
\beq
{D \over D_H} \rightarrow {2\over 1+3w}.
\slabel{soverdh1}
\eeq
Thus, scales  larger than the Hubble horizon remain
out of casual contact until the Hubble horizon grows to be 
comparable to them.

It follows that a patch of space whose density or other matter
variables are different from those of the surrounding space, and whose
size is larger than the Hubble distance during inflation, will evolve
like a separate homogenous Universe. It will continue to do so until
the Hubble distance becomes comparable to the patch size. The
difference between a matter variable in the patch and outside the
patch can be evaluated by solving the background equations for the
background space-time and those for the patch and then subtracting the
difference between the two. The coordinate freedom of the time surfaces
on which to match the patch and background is the same as the usual
gauge freedom in cosmological perturbations \cite{Bar80}. If the
coordinate system is chosen so that the patch and the background have
the same scale factor or equivalently the same efoldings, then, under
reasonable assumptions, the difference between the matter variables in
the patch and in the background space-time is
the same as the perturbation in the flat gauge \cite{sepuni}. In a
homogenous space-time, the state of the Universe at any time can be
totally specified in terms of the degrees of freedom such as the
different fluids' densities and pressures and the values and time
derivatives of the scalar fields. 
Thus, the large scale, flat gauge perturbation, $\delta f$, of a
function, $f$, of the matter 
degrees of 
freedom, $\phi_i$,
can be evaluated as
\beq
\delta f = \sum_i {\partial f \over \partial \phi_i|_{N=0}} \delta \phi_i|_{N=0}.
\slabel{sepunieq}
\eeq
This equation summarizes the separate Universe approach and will be
used in evaluating  matter perturbations in the rest of the paper.

\section{Quintessence Background Equations}
The Klein-Gordon equation for the quintessence, $Q$, is given by
\beq
{\partial^2 Q \over \partial t^2}+ 3H{\partial Q\over \partial t} +
{\partial V \over \partial Q}=0 
\slabel{KG}
\eeq
where $V$ is the quintessence potential.
We can express the Klein-Gordon equation (\eq{KG}) in terms of $N$, by
using the Friedmann equation (\eq{friedmann}) and the acceleration
equation (\eq{acceleration}), as
\beq
Q'' + {3 \over 2}(1-w)Q' + {1\over H^2} {\partial V \over \partial Q}=0
\slabel{KG1}
\eeq
where the prime indicates differentiation with respect to the number
of efolds, $N$, and $w$ is the total equation of state parameter (\eq{w}), including the
effects of any other matter present.
If the value of $Q$ does not change much during the period of
interest, then its potential may be approximated by a constant plus a
linear term 
\beq
V \approx V_* \left( 1 + {Q-Q_* \over M_p} \sqrt{2\epsilon_*} \right)
\slabel{V}
\eeq
where $V_*$ and $\epsilon_*$ are constants and correspond to the
potential and first slow roll parameter, at the point about which the
expansion is taken, $Q_*$, respectively. A class of potentials that do
not satisfy this criteria in general are the `tracking' potentials
 which require \cite{SteWanZla99}
\beq
{1\over V_Q}{\partial^2 V_Q \over \partial Q^2 }\left( {1\over V_Q}{\partial V_Q \over
   \partial Q} \right)^{-2} \ge 1.
 \eeq
In the tracking regime, the solutions are insensitive to changes in the initial
conditions of the quintessence. It follows, from the discussion
in Sec.~\ref{sepunisec}, that a patch with different initial conditions
for the quintessence, will quickly approach the
background solution for the quintessence, 
and so the perturbations will be suppressed \cite{tracksup,BarCorLidMal04,GorHu04}. 

Substituting the linear form of the potential (\eq{V}) into the
Klein-Gordon equation (\eq{KG1}) and using the Friedmann equation
(\eq{friedmann1}) gives
%\begin{eqnarray}
\beq
%&
Q'' + {3\over 2} (1-w) Q'
%&  \nonumber \\
%&
 + 3\sqrt{2} \sqrt{\epsilon_*}{M_p V_* \over
    \rho_0} {\rm e}^{3(1+w)N} = 0.
%&
\slabel{KG2}
%\end{eqnarray}
\eeq
Assuming the total equation of state parameter ($w$) is a constant,
less than one,
and denoting the values of $Q$ and its derivative, at $N=0$, by $Q_0$
and $Q_0'$ respectively, the
solution to the Klein-Gordon equation (\eq{KG2}) for large $N$ is 
\beq
Q = Q_0 + {2 \over 3 (1-w)}Q_0'
- {2\sqrt{2}\sqrt{\epsilon_*} \over
  3(3+4w+w^2)} {M_p V_* \over \rho_0 {\rm e}^{-3(1+w)N}}.
\slabel{Q}
\eeq
As $\epsilon_*<1$ is needed for the quintessence to cause acceleration, and 
$V \lesssim \rho$ today, it follows that the last term in the above
equation will be negligible until a redshift of about one. It
follows that, the quintessence field is frozen until about that point.
This does not apply to  the case where the field starts rolling
in the early Universe due to a steep potential and then rolls into an
area of parameter space where the potential is shallow.

The energy density of the quintessence is 
\beq
\rho_Q \equiv V + {1\over 2}\left({\partial Q \over\partial t}\right)^2 =
V+{1\over 2} H^2 Q'^2.
\slabel{rhoQdef} 
\eeq
Substituting the solution for $Q$ (\eq{Q}) and the linear potential
(\eq{V}) into the above equation gives
%\begin{eqnarray}
\beq
%&
\rho_Q = V|_{Q=Q_0} 
% & \nonumber \\ &
 +V_* \left( {2\sqrt{2\epsilon_*}\over
  3(1-w)}{Q_0' \over M_p} 
 - {8\epsilon_*\over 3(1+w)(3+w)^2} {V_*
  \over \rho_0 {\rm e}^{-3(1+w)N}} \right).
%&
\slabel{rhoQ}
%\end{eqnarray}
\eeq
This shows that the density will also be frozen until about $z\sim 1$.

\section{Perturbations}
Using the separate Universe approach (\eq{sepunieq}), the density
perturbation of the quintessence, in the flat gauge and on large
scales, can be written as  
\beq
\delta \rho_Q = {\partial \rho_Q \over \partial Q_0} \delta Q_0 +
       {\partial \rho_Q \over \partial Q_0'} \delta Q_0' + {\partial
	 \rho_Q \over \partial \rho_0} \delta \rho_0.
\slabel{deltarhoQdef} 
\eeq
Substituting the result for the quintessence density (\eq{rhoQ})
into the above equation and using the background density equation
(\eq{density}) 
gives  
%\begin{eqnarray}
\beq
%&
\delta \rho_Q = V_* \left( \sqrt{2\epsilon_*} {\delta Q_0 \over M_p
 } \right. 
%& \nonumber \\&
 \left. + {2\sqrt{2\epsilon*} \over 3(1-w)}{\delta Q_0' \over M_p} + {8
  \epsilon_* \over 3(1+w)(3+w)^2} {V_* \over \rho} {\delta \rho_0
  \over \rho_0} \right).
%&
\slabel{deltarhoQ}   
%\end{eqnarray}
\eeq
As can be seen from the above equation, the quintessence will acquire
large scale density perturbations even if it is initially
homogenous. This is in contrast to two fluid components which, as can
be seen from the density equation (\eq{density}), only depend on their
own initial perturbation, and so if one is initially homogenous it will
remain homogenous on large scales. The reason why the scalar field
acquires a perturbation is from the coupling to the Hubble parameter
in the Klein-Gordon equation (\eq{KG1}). In the case of a fluid, this
coupling is lost in the continuity equation (\eq{continuity}) when the
time parameter is converted to efolds.

Using the solution for the quintessence density (\eq{rhoQ}) and density
 perturbation (\eq{deltarhoQ}), it can be shown that 
\beq
{\delta \rho_Q|_{\delta Q_0=0,\delta Q_0'=0} \over \rho_Q'} =
{\delta  \rho_0 \over \rho_0'}.  
\eeq
This shows that the perturbation in $\rho_Q$ becomes {\em adiabatic}
if the scalar field is initially unperturbed. 

The magnitude of the adiabatic perturbation can be evaluated from the
solution for the perturbation (\eq{deltarhoQ}) and background density
(\eq{rhoQ}) to give, assuming $\epsilon_*  V_* / \rho \ll 1$,
\beq
{\delta \rho_Q|_{\delta Q_0=0,\delta Q_0'=0}\over \rho_Q |_{Q=Q_*,Q_0'=0}}
= {8 \epsilon_* \over 3(1+w)(3+w)^2} {\rho_Q \over \rho} {\delta
  \rho_0 \over \rho_0}.
\slabel{deltarhoad}
\eeq 
The value of the perturbation in the matter variable, $\rho_0$, can be
evaluated by its relation to the measured value of the curvature
perturbation on constant density hyper-surfaces \cite{BarSteTur83}. In the
flat gauge, this is given by \cite{sepuni}
\beq
\zeta \equiv {\delta \rho_0 \over 3(1+w)\rho_0}.
\slabel{zeta} 
\eeq
The WMAP CMB measurements  constrain $\zeta$ \cite{Spe03} and so give
the variance of the density 
perturbation to be 
\beq
{\left<\delta \rho_0^2\right>^{1/2} \over {3(1+w)\rho_0} }\approx 5\times 10^{-5}.
\slabel{deltarhomeas}
\eeq
Substituting the above equation into the expression for the adiabatic
perturbation (\eq{deltarhoad}) gives
\beq
{\left<\delta \rho_Q^2\right>^{1/2}|_{\delta Q_0=0,\delta Q_0'=0}\over \rho_Q
  |_{Q=Q_*,Q_0'=0}} < 4 \times 10^{-5}.
\slabel{deltarhoadlim}
\eeq
Using the WMAP data, the quintessence density perturbation is measured
to be \cite{GorHu04}
\beq
 {\left<\delta \rho_Q^2\right>^{1/2} \over \rho_Q} = (8 \pm 4) \times
 10^{-4}
 \slabel{deltarhoqmeas}
 \eeq
at the 68$\%$ confidence interval.
It follows  that the adiabatic perturbation (\eq{deltarhoadlim}) is  small in comparison to this value. During inflation, if the quintessence is a light field it
will acquire the usual large scale perturbations which depend on the
Hubble parameter,
\beq
\left<\delta Q_0^2\right>^{1/2} = {H_{\rm inf} \over 2\pi}. 
\eeq
As gravitational waves have not been detected in the WMAP data, this
puts an upper limit on this quantity \cite{Pei03}
\beq
\left<\delta Q_0^2\right>^{1/2} < 9 \times 10^{-6} M_p.
\eeq
Using the above equation in the solution for the quintessence density
perturbation and background value (\eqs{deltarhoQ}{rhoQ}) gives the
non-adiabatic part, from inflation, of the quintessence density
perturbation as 
\beq
{\left<\delta \rho_Q^2\right>^{1/2} |_{\delta Q_0'=0,\rho_Q \ll \rho_0} \over
  \rho_Q|_{Q_0=Q_*,\delta Q_0'=0, \rho_Q \ll \rho_0}} < 10^{-5}
\eeq
which is also too small compared to the observationally preferred
value (\eq{deltarhoqmeas}).

\section{Conclusions}
Using the separate Universe approach, I have investigated the perturbations
in the quintessence by perturbing the background solutions. It was
shown how the quintessence perturbation and background value can be
frozen from inflation until about dark energy domination. Also the way
in which a homogenous quintessence field acquires an adiabatic
perturbation was illuminated.

Both the adiabatic perturbation and the non-adiabatic perturbation in
the quintessence from inflation where shown to be smaller than the
value preferred by observations. 
This problem was originally 
identified in references \cite{MorTak03,GorHu04} using the Newtonian and
comoving curvature gauges.

In this paper, the adiabatic perturbation was shown
to be determined by the perturbation in the
non-relativistic matter and so fixed by observations. However, the
non-adiabatic perturbation from $\delta Q_0$ can be made large by
making the quintessence density more sensitive to the initial value of
$Q_0$. This requires some evolution in $\delta Q$.  A mechanism for
amplifying the quintessence 
perturbation is given in reference \cite{GorWan05}.    

\smallskip{\it Acknowledgments:}
I thank Wayne Hu  and  Dragan Huterer for useful discussions. 
Also, I  was supported by the KICP under NSF PHY-0114422.

\end{document}